\documentclass[aas2pp4]{aastex}
\input{psfig} 

\slugcomment{Version 2}

\shorttitle{K-band Spectroscopy of Clusters in NGC 4038/4039}
\shortauthors{Mengel et al.}

\begin{document}

\title{K-Band Spectroscopy of Compact Star Clusters in NGC 4038/4039}

\author{Sabine Mengel, Matthew D. Lehnert, Niranjan Thatte, 
Lowell E. Tacconi-Garman and Reinhard Genzel}
\affil{Max-Planck-Institut f\"ur extraterrestrische Physik, Postfach 1312, D-85741, 
Garching, Germany}
\email{mengel, mlehnert, thatte, lowell, genzel@mpe.mpg.de}

\begin{abstract}

Integral field spectroscopy in the K-band (1.9-2.4$\mu$m) was performed
on four IR-bright star clusters and the two nuclei in NGC 4038/4039 (``The
Antennae'').  Two of the clusters are located in the overlap region of
the two galaxies, and together comprise $\approx$25\% of the total
15$\mu$m and $\approx$10\% of the total 4.8 GHz emission from this pair
of merging galaxies.  The other two clusters, each of them spatially
resolved into two components, are located in the northern galaxy, one in
the western and one in the eastern loop of blue clusters.  Comparing our
analysis of Br$\gamma$, CO band-heads, He I $\lambda$2.058$\mu$m,
H$\alpha$ (from archival HST data), and V$-$K colors with stellar
population synthesis models indicates that the clusters are extincted
(A$_V$$\approx 0.7 - 4.3$ mags) and young, displaying a significant age
spread (4-13 Myrs).  The starbursts in the nuclei are much older
(65 Myrs), with the nucleus of NGC 4038 displaying a region of recent
star formation northward of its K-band peak.  Using our derived age
estimates and assuming the parameters of the IMF (Salpeter slope, upper
mass cut-off of 100 M$_{\sun}$, Miller-Scalo between 1 M$_{\sun}$ and
0.1 M$_{\sun}$), we find that the clusters have masses between 0.5 and 5
$\times$ 10$^6$ M$_{\sun}$.

\end{abstract}


\section{Introduction}

Arguably, one of the most fascinating recent astrophysical discoveries is
the rich population of young luminous blue ``globular-like'' clusters in
strongly interacting and merging systems \citep[][hereafter WS95]{Hetal92,
Wetal93, Wetal97, WS95}.  These clusters all appear much younger ($<$
100 Myrs) than both the old globular cluster system and underlying
background population and have dynamical masses \citep[M$\sim$few $\times$
10$^5$ M$_{\sun}$ -- in a few cases --][]{HF96} and effective radii
very similar to the globulars (r$_e$$\sim$10 pc -- WS95).  The formation
of such clusters appears to be generic to the merger process and
is likely the result of the high pressures that are induced during the
merger process \citep[e.g.,][]{EE97}.

The star formation activity in NGC 4038/4039 -- the ``Antennae'' -- has
been known from studies at various wavelengths, for example from
H$\alpha$ spectroscopy \citep{Ru70} or the very blue colors observed in
the HST images (WS95).  Its relative proximity (19 Mpc assuming
H$_0$=75km s$^{-1}$Mpc$^{-1}$) makes it the ideal candidate for the
study of merger induced star formation on scales of single star clusters
and has allowed for the detection and investigation of a very large
number ($\approx$800) of compact clusters \citep[WS95]{W99}.  The
heavily extincted overlap region was revealed to be the most active star
formation zone by radio observations \citep{HvdH86} and imaging and
spectroscopy performed with ISO \citep{MVCSGTCMD98, Ketal96, Fetal96}.
The overlap region is, however, relatively inconspicuous at optical
wavelengths.  The ISO results suggest very young ages for these clusters
and the presence of high mass (50-60 M$_{\sun}$) stars.  Interestingly,
a significant fraction of the total bolometric flux from the Antennae
appears to originate in two clusters eastward of the NGC 4039 nucleus
(numbers 86/87/88/89/90 in WS95), building up an optically bright knot,
and number 80 in WS95, which is optically extincted and declared a
``very red object''.  These two regions, which are the highest surface
brightness features in the mid-infrared, contribute $\approx$25\% of the
total flux at 15$\mu$m \citep{MVCSGTCMD98} and $\approx$20\% of the
total emission from discrete radio sources in NGC 4038/4039
\citep{HvdH86}.  These large fractions suggest that these two regions
must constitute a significant portion of the very recent or on-going
star-formation within the Antennae.  Understanding their properties
within the perspective of the overall star formation history of the
Antennae is of prime importance to understanding the bolometric output
of these merging galaxies.  We discuss the properties of these regions
as determined through IR integral field spectroscopy in combination with
data available from the literature.

\section{Cluster Selection and Observations}

\begin{figure*}[ht]
\figurenum{1}
\psfig{file=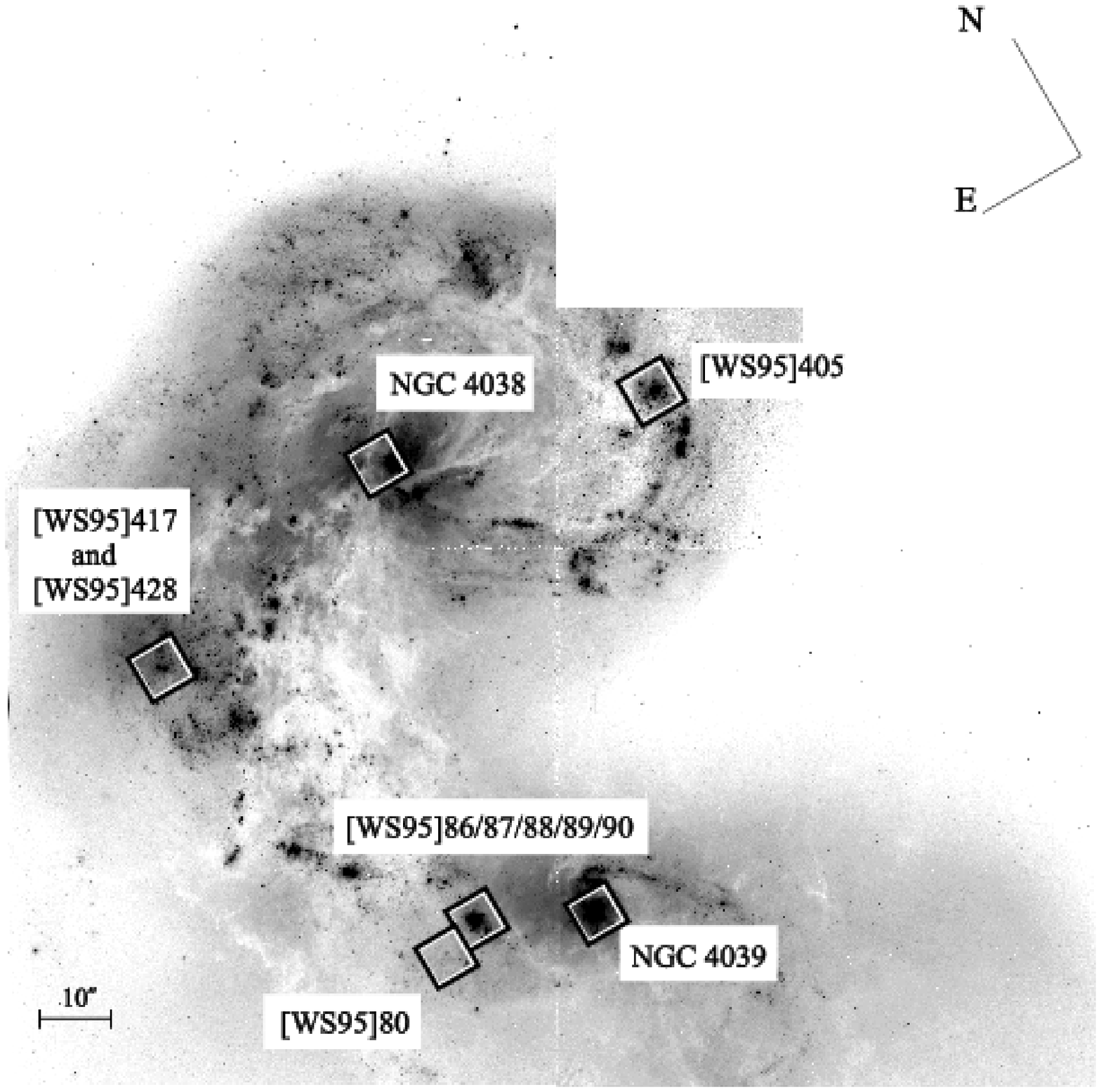,width=16.5cm}
\caption{The HST V-band image \citep{W99} displayed in a logarithmic
grayscale with the 3D-fields and their numbering (referring to
\citet{WS95}) as used in this paper indicated.  The overall morphology
is that of two colliding gas-rich spiral galaxies (likely Sc type).
Eastward of the line connecting the two nuclei lies the so-called
``interaction zone'' or ``overlap region'', which is heavily extincted
in the optical.  The size scale is indicated in the image.}
\end{figure*}

Models of the merger of two spiral galaxies (like that currently being
observed in NGC4038/4039) predict multiple bursts of star formation during
the merger event \citep{MH96}.  Depending on the stage of the merger,
these bursts can be in the region of overlap of the two galaxy disks, in
the tidally shocked regions of the outer disks of each galaxy, or as
significant quantities of gas fall in towards the two nuclei, in the
nuclei of each galaxy.  We observed the two most prominent clusters
within the interaction zone, two other IR bright clusters in the outer
arms of NGC 4038/4039 (likely to be tidally shocked regions), and the two
nuclei.  Such a selection would allow us (at a minimum) to probe the
various regions of star-formation during the merger process within the
inevitable constraint of only having a limited amount of observing time.
While only observing six regions will not allow us to investigate the
properties of the whole ensemble of compact clusters in the Antennae,
our sample allows us to crudely investigate the possible range of ages
across the Antennae.  Figure 1 shows the two clusters in the interaction
zone ([WS95]86 and [WS95]80), the two nuclei (NGC 4038 and NGC 4039) and
two clusters in the outer loops ([WS95]417/[WS95]428 in the eastern and
[WS95]405 in the western), observed with the 3D integral field
spectrometer \citep{Wetal96, Ketal95} in the K band.


For the observations, 3D was combined with the tip-tilt-guider ROGUE
\citep{T95} on the Anglo-Australian Telescope in February, 1998 and
April, 1999.  We used a K-band grism with a resolution of $\lambda /
\Delta\lambda = 1000$ and a 256x256 NICMOS 3 array.  256 spectra are
obtained simultaneously, arranged in a 16x16 pixel field on the sky.
The detector integration time per exposure was 100 seconds.  The
on-source exposures were interleaved with off-source (offset 60\arcsec E
of each cluster) exposures of identical integration time for the sky
subtraction.  To remove telluric features, the A2V star HD106819 was
observed roughly once every hour.  In order to Nyquist sample the
spectra, we dithered in the spectral direction by half a pixel using a
piezo-driven mirror.  The spatial pixel scale of 0\farcs4 provided a
field of view of 6\farcs4x6\farcs4.  

The conditions were generally not photometric and flux calibration
achieved via aperture photometry available from the literature \citep{BS92} in
combination with Ks-band imaging (see also next section).  The seeing during the
observations of each individual region only showed moderate variation
($<$0\farcs2) but was significantly different for different objects.
The values for each observation are discussed in section \ref{results}.
The total on-source integration time was 6200s on the first field
(hereafter abbreviated to [WS95]86), which is the optically bright star
cluster, corresponding to the numbers 86/87/88/89/90 in WS95; 4200s
on-source time for the second field (hereafter [WS95]80), which corresponds
to the ``very red object'' number 80 in WS95; 3120s on-source time for
the field in the eastern loop (numbers 417 and 428 in WS95); 3840s for
the targets in the western loop (WS95 - number 405 and several fainter
objects, hereafter [WS95]405).  The nuclei of NGC 4038 and NGC 4039 were
observed for 3360s and 2160s respectively.  Due to a piezo failure, the
spectrum of the nucleus of NGC 4039 is not Nyquist sampled.

\section{Data Reduction and Analysis}

Data reduction was performed using the 3D data reduction routines
developed at MPE within the GIPSY \citep{vdh92} data reduction
environment.  Single frames were dark subtracted, linearized and sky
subtracted using the sky frame that was closest chronologically and had
the right spectral dithering position.  The wavelength calibration was
accomplished by exposures of a neon discharge lamp.  The re-binning of
the spectrally interleaved data was performed onto a 600 pixel linear
wavelength axis.  Bad pixels and cosmic rays were removed from the
16x16x600 data cube, either by interpolation or by masking out.  After
spatial re-binning onto a 64x64 pixel grid, the single cubes were
stacked by centering each on its K-band continuum peak or the Br$\gamma$
peak for [WS95]86 and [WS95]80, respectively.  Telluric features were
removed by division by the atmospheric transmission profile obtained
from the observations of the A2V star.  The Br$\gamma$ absorption line
in the calibrator spectrum was interpolated linearly between the
adjacent continuum.  In the resulting cluster spectra, only the NGC 4038
nucleus spectrum shows a little remaining Br$\gamma$ emission from the
calibrator, which is not a worry, because it is at zero redshift and
does not contaminate the Br$\gamma$ emission from the object.

Object spectra were extracted using a square aperture centered on the
continuum peak and sized to maximize the SNR of the resulting spectrum.
The individual apertures and their sizes are shown in Figure 2.  For
fields which showed significant offsets between continuum and Br$\gamma$
line emission peaks ([WS95]405 and the nucleus of NGC 4038), the
aperture sizes were chosen so as to minimize contamination between the
continuum and line emitting regions.  The spectra were flux calibrated
using the K$_s$-magnitudes from our flux calibrated SOFI image, obtained
in May 99 at the NTT.  The extracted spectra were normalized by dividing
by the continuum value, estimated using a linear fit.  From this we
estimated the equivalent widths W$_{\lambda}$ of the Br$\gamma$- and the
He I-lines and the CO-band-heads.  For the creation of linemaps, the
linear fit was performed for each spatial pixel and subtracted.  The
spectra of all the regions (divided into subregions for some of the
fields) are shown in Fig. 2, together with the K$_s$-band images.
Overlaid are the contours of Br$\gamma$ or the H$_2$(1-0)S(1)-line
(labelled in Fig. 2).

\section{Results} \label{results}

\begin{figure*}[ht]
\begin{center}
\figurenum{2}
\psfig{file=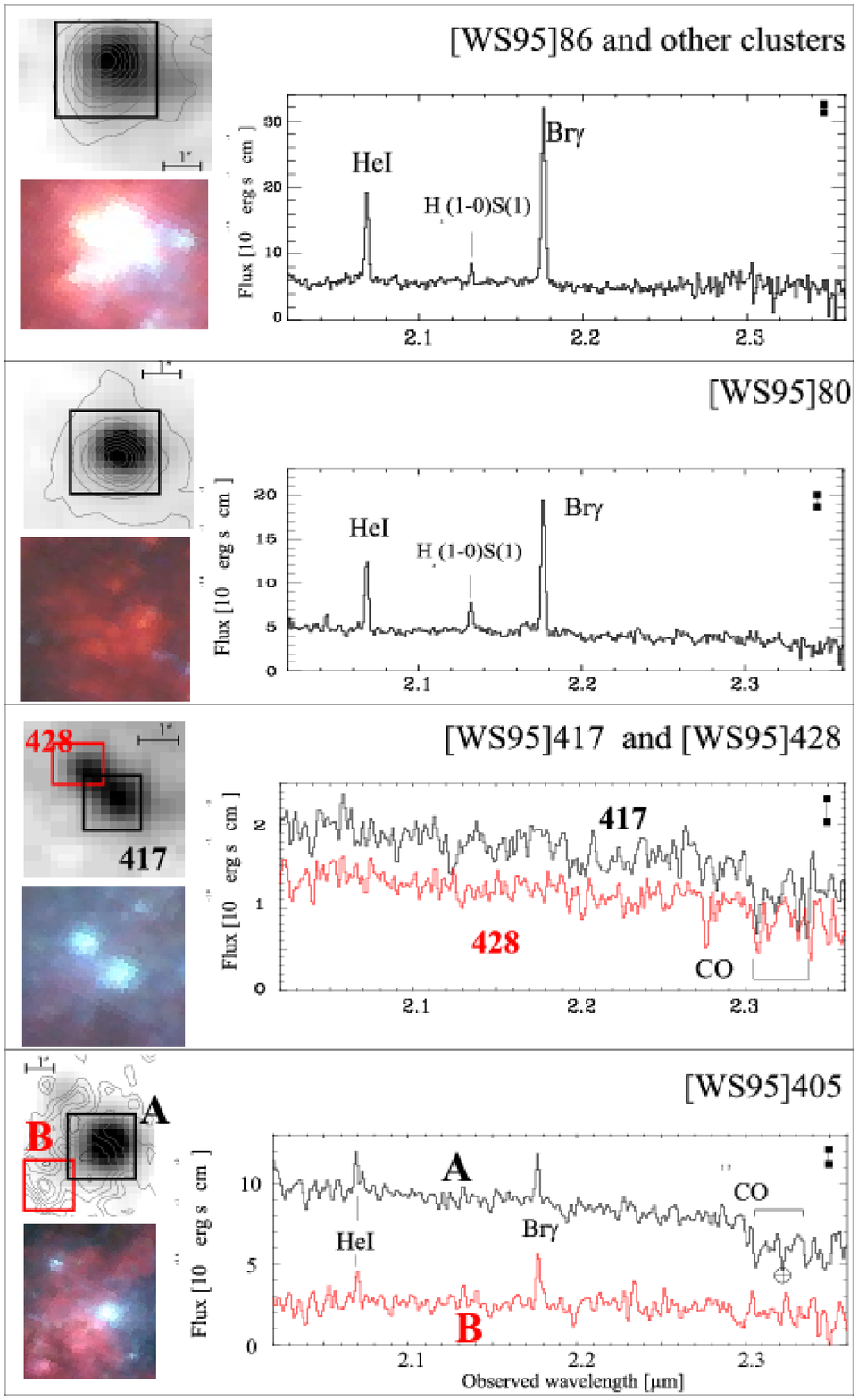,width=13cm}
\end{center}
\end{figure*}
\begin{figure*}[ht]
\begin{center}
\figurenum{2}
\psfig{file=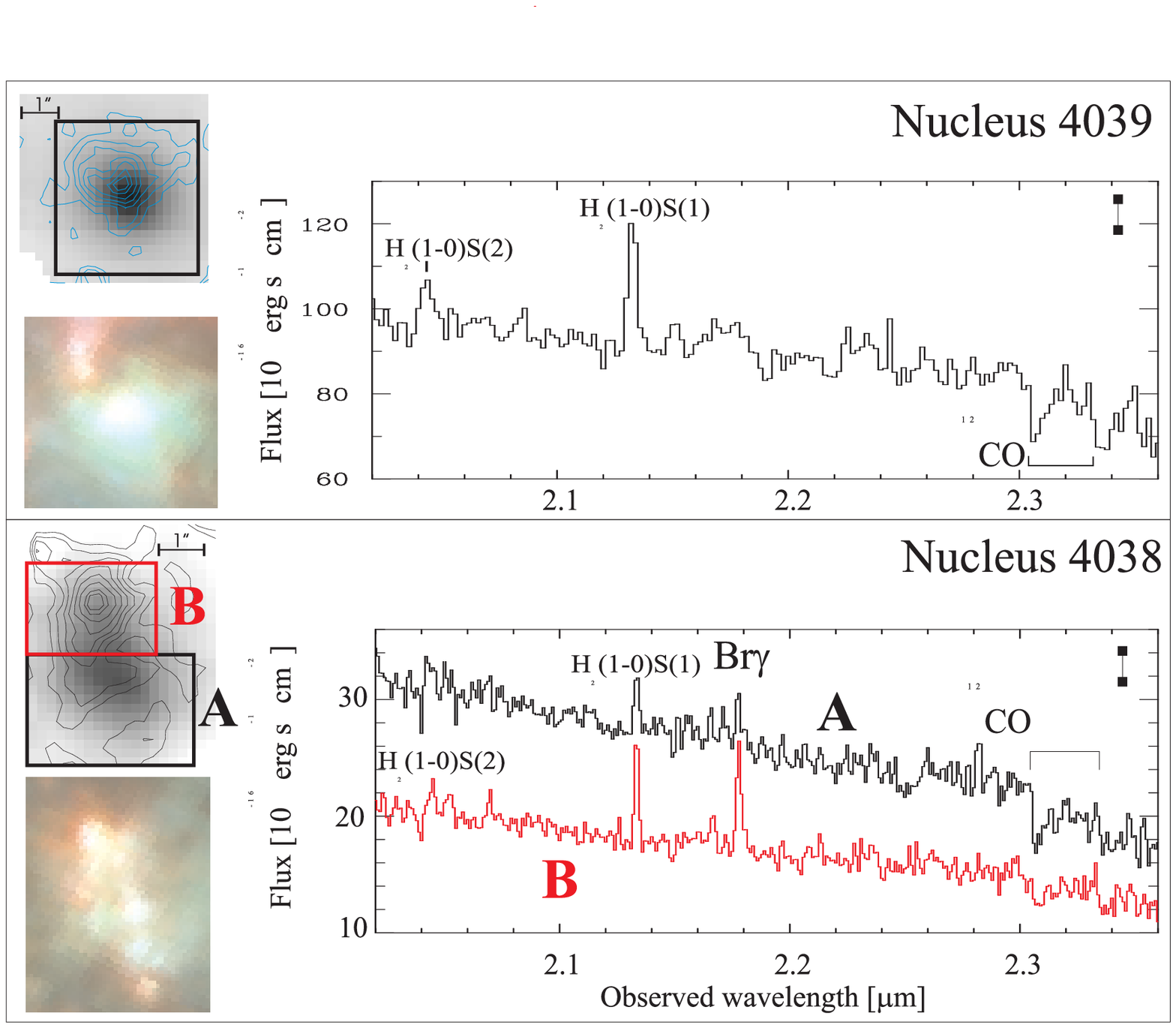,width=15.5cm}
\caption{K-band images of the observed fields with Br$\gamma$ or H$_2$
contours, K-band spectra and the corresponding regions in color
composite images taken from W99 (blue:  U+B, green:  V, red:
H$\alpha$).  The greyscale images in the left column (top) are created
from the 3D data cubes by summation of the corresponding channels of the
data cubes, continuum subtracted in the case of the emission line maps.
They show the Br$\gamma$ distribution in [WS95]86, [WS95]80,
[WS95]417\&428, [WS95]405 and nucleus of NGC 4038, and the
H$_2$(1-0)S(1) distribution in the nucleus of NGC 4039.  The highest
contours represent 95\% of the maximum flux, then from 90\% continuing
in steps of 10\% downward.  Maximum fluxes are 2.0 ([WS95]86), 1.7
([WS95]80), 0.47 ([WS95]405), 0.59 (NGC 4039 nucleus) and 0.46 (NGC 4038
nucleus) $\times 10^{-16}$ erg s$^{-1}$ cm$^{-2}$.  The K magnitudes of
the fields are given in Table 1.  The spectra were extracted from the
regions indicated in the K-band images.  The fluxes are given in units
of $10^{-16}$ erg s$^{-1}$ cm$^{-2}$ versus the wavelength in $\mu$m.
The noise level is indicated in the top right corner of each spectrum.
The images in the left column (bottom) show the corresponding regions of
the HST multi-color image (W99) with U+B in the blue, V in the red and
H$\alpha$ in the red display.  The scale is given as 1\arcsec\/ bars
(the pixel size in all 3D images is 0\farcs2 x 0\farcs2).  At the
distance of the Antennae 1\arcsec\/ corresponds to 93 pc.  N is up, E to
the left in each image.}

\end{center} 
\end{figure*}

Results obtained from our data, incorporating results from publicly
available HST data \citep[H$\alpha$ and V-band images --][]{W99}, are
listed in Table 1.  [WS95]86 was slightly resolved while [WS95]80 was
unresolved (both combined data sets had a final FWHM$\sim$1\farcs0; Fig.
2).  The FWHM of the seeing was $\sim$0\farcs8 during [WS95]417\&428
observations, and the two clusters are marginally resolved into single
components.  During [WS95]405 observations, the seeing was poor
(1\farcs6), but to the NE of the bright cluster, a fainter second
cluster is visible.  The tabulated values for [WS95]405-A and B refer to
the bright cluster (A) and to the location of the Br$\gamma$ peak (B).

We used the measured equivalent width of Br$\gamma$, W$_{Br\gamma}$, and
CO 2.29$\mu$m band-head, W$_{CO}$ (Table 1), to determine the ages of
the starbursts using stellar synthesis models \citep{Letal99, KSA98}.
See plots in \citet{Letal99} (their Figures 89 and 101), showing the 
evolution of
these parameters with burst age.
Assuming an upper mass cutoff M$_{upper}$=100M$_{\sun}$, Salpeter IMF
slope and an instantaneous burst at solar metallicity, we derive ages of 3.7$^{+1.0}_{-0.4}
\times 10^6$yr for [WS95]86 and 5.5$^{+0.7}_{-0.8} \times 10^6$yr for
[WS95]80.  \citet{GNIR00} derive a value of $\sim$4 Myrs for [WS95]80,
assuming the same starburst parameters, in reasonable agreement with our
value.  The ages of the two clusters in [WS95]417\&428 are not very well
constrained, due to the low S/N of the data and the ambiguous value of
the W$_{CO}$.  
However, the values of W$_{CO}$ are sufficiently large to exclude
additional errors introduced by the non-inclusion of the thermally
pulsing asymptotic giant branch phase in the Geneva stellar evolution
tracks that were incorporated in \citet{Letal99} \citep{OO00}.
Age estimates of [WS95]417\&428 lie between 8.5 and 12.8
Myrs old.  For the [WS95]405 spectra, it is not possible to get a
consistent model age with the presence of both strong CO-band-heads and
Br$\gamma$ emission with an instantaneous burst.  But given the size of
the cluster \citep[R$_{eff}$=12.3pc,][]{W99} it is plausible to assume
that the star-formation has a finite duration of a few Myrs
\citep{EfEl98}.  Other possibilities that might explain why these
clusters have both CO absorption and Br$\gamma$ emission are shock
heating of the gas due to supernova explosions, the seeing broadened
Br$\gamma$ emission from the region to the S-E of [WS95]405-A which
contaminates the cluster spectrum, supersolar metallicity, and possible 
gaps in our knowledge of
stellar evolution and model atmospheres.  It is interesting to compare
the 3D image and spectra of a region outside both [WS95]405 clusters
with the multi-color HST image, since our data show a Br$\gamma$ peak to
the SE of the K-band continuum peak of cluster A.  This coincides with a
dust filament obvious in the HST image, and it could be the site of more
recent star formation or of shock excitation in a more dusty
environment.

A detailed discussion of the nuclear spectra is beyond the scope of the
current paper.  Here we only wish to emphasize how the spectra of the
two nuclei differ from the young compact clusters.  The W$_{CO}$ of the
two nuclei suggest the same possible age range for both, 65$\pm$15 Myrs,
considerably older than the ages estimated for the prominent young
clusters.  The northern nucleus also shows evidence for a more recent
starburst to the north of the K-band peak, but it is difficult to
estimate its age due to the underlying continuum contribution by the
older background nuclear stellar population.  But the presence of
Br$\gamma$ emission (no evidence for AGN activity was seen in other
observations so far) requires the presence of O or B stars or recent
shock excitation, thus implying that it is younger than $\sim$10-20
Myrs.  Under the assumption that there is no intrinsic gradient in
W$_{CO}$ across the NGC 4038 nucleus, meaning that the observed lower
W$_{CO}$ at the location of the Br$\gamma$ peak can be attributed to the
additional continuum from the young stellar population, we estimate the
young star light continuum to contribute about 1/5 of the total
continuum, which increases the observed W$_{Br\gamma}$ to 58\AA.  This then
implies an age of the starburst of around 6 Myrs, consistent with
the effective temperature of 33000K derived from the low ratio of He
I/Br$\gamma$ of 0.17.

The age uncertainties given in Table 1 only take into account
uncertainties in the signal-to-noise of the data and do not take into
account the systematic effects of selecting different IMF slopes, upper
and lower mass cutoffs, abundances, or uncertainties in the models
themselves.

The extinction was derived by comparing the theoretical ratio of
H$\alpha$/Br$\gamma$ \citep[case B recombination ratio assuming
T$_e$=10000 K from][]{SH95} with the observed ratio.  Using the
extinction curve of \citet{D89} for a foreground screen model, we
derived A$_V$ of 1.4$\pm$0.2 and 4.3$\pm$0.2 for [WS95]86 and [WS95]80,
respectively.  They agree within the errors with the values we derived
from the V-K color excess with the intrinsic color derived from the
models for the respective cluster ages (A$_V$ = 1.6$^{+0.6}_{-1.0}$ and
3.7$^{+0.6}_{-1.0}$).  For [WS95]417\&428 and [WS95]405, we used the V-K
colors not to confirm the extinction, but rather to further constrain
the age.

The ratio of He I $\lambda$2.058$\mu$m/Br$\gamma$ is not an unambiguous
tracer of the effective temperature of the highest mass stars present in
a stellar population \citep{S93,Natasha99}, but it was determined to
compare with the ISO results at higher spatial resolution.  The effective
temperatures derived from the ratios of He I/Br$\gamma$=0.47 and 0.5 are
36,000 and 38,000K ([WS95]86 and [WS95]80, respectively), which
corresponds to masses of roughly 28 M$_{\sun}$ and is below the 44,000K
(50-60 M$_{\sun}$) derived from mid-infrared line ratios by
\citet{Ketal96}.  Given the sensitivity of the He I $\lambda$2.058$\mu$m
line to nebular parameters \citep{S93} and the large aperture
(14''$\times$27'') used for the ISO measurement, some disagreement
between our estimate and that obtained using ISO data for the
temperature of the hottest stars is not surprising.  Moreover, the
relationship between temperature of the hottest stars and He
I/Br$\gamma$ is not monotonic and thus our estimates for the mass should
be considered lower limits.  These upper mass limits are consistent with
our estimated ages of the [WS95]80 cluster, while the younger age of
[WS95]86 would allow for the detection of higher mass stars.  A
qualitative comparison of the derived effective temperatures/ages of all
fields agrees with the gradient in the $[$Ne III$]$/$[$Ne II$]$ observed
by \citet{Vetal96}.

From the extinction corrected Br$\gamma$ flux we can estimate the number
of Lyman continuum photons, N$_{Lyc}$.  For this calculation, we used
the extinction corrected Br$\gamma$ flux, a distance of 19.2 Mpc for the
Antennae, and recombination coefficients from \citet{HS87} and
\citet{SH95}.  We find that N$_{Lyc}$=3.1$\pm0.3\times$10$^{52}$
s$^{-1}$ and 1.9$\pm0.4\times$10$^{52}$ s$^{-1}$ for [WS95]86 and
[WS95]80 respectively.  This somewhat lower than the value derived from
the thermal radio continuum flux density estimated by \citet{HvdH86},
which results in N$_{Lyc}=1.3\pm0.7\times10^{53}$ s$^{-1}$ for the sum
of [WS95]86 and [WS95]80.  Given that our apertures are smaller, the
difficulty in estimating the thermal contribution to the total radio
emission, and that there is some diffuse radio emission, it is not
surprising that there is a difference between the two estimates.  It is
also significantly less than the value of 1.0$\times10^{53}$ photons
s$^{-1}$ derived by \citet{GNIR00}.  Both our observed flux and
estimated extinction are lower by factors of $\sim2$ and both of these
differences account for the discrepancy between our respective estimates
of the Lyman continuum flux.  However, \citet{GNIR00} caution that the
extinction value they derived (A$_K$ = 1.1mag) should be considered an
upper limit.

Comparing our results (M$_K$ (0) and N$_{Lyc}$) with the models of
\citet[assuming a Salpeter IMF slope, M$_{upper}$=100 M$_{\sun}$, high
mass stars have already evolved off the main sequence and are therefore
not observed, and M$_{lower}$=1 M$_{\sun}$]{Letal99} for the age
determined for each cluster in the previous section, the total mass of
each cluster would be 1.6$^{+1.2}_{-0.2}$ and
3.0$^{+3.6}_{-0.7}\times$10$^{6}$ M$_{\sun}$ for [WS95]86 and [WS95]80
respectively.  A lower mass limit equal to 0.1 M$_{\sun}$ (using
Miller-Scalo slope for the lower mass end of the IMF) would imply masses
a factor of 1.6 higher.  For [WS95]417\&428 and [WS95]405, we crudely
estimated their masses using only their extinction-corrected
K-magnitudes compared to the magnitudes expected from our derived ages.
Their masses are lower than the masses of [WS95]86 and [WS95]80, and lie
between 0.3 and 1.1 $\times 10^6$ M$_\sun$.

This places each of the very massive [WS95]86 and [WS95]80 clusters into
the same mass regime as M82A, which produces roughly the same amount of
N$_{Lyc}$, and has the same burst age \citep{Natasha99}.  Their masses
are more than a factor of ten higher than the average mass of a globular
cluster \citep{MSS91}.  Even if they are expected to lose 60\% of their
mass over a Hubble time \citep{CW90}, they will still represent the
top-end of the globular cluster mass function.  The [WS95]417\&428 and
[WS95]405 clusters have masses which are more moderate and comparable to
the average globular cluster mass.  This mass range in the observed
clusters is not surprising given that those two fields contribute a
substantial amount of the bolometric luminosity of the Antennae.  But a
more accurate mass estimate of a larger fraction of the population of
young star clusters is necessary to judge if the young clusters could
evolve to form a part of the globular cluster population of an
elliptical galaxy.

\section{Summary and Conclusions}

We performed integral field spectroscopy in the K-band (1.9-2.4$\mu$m)
on four star clusters and the nuclei in NGC 4038/4039 in order to derive
the starburst properties and their variations with the location in the
merger.  Our analysis of Br$\gamma$, CO band-heads, He I
$\lambda$2.058$\mu$m/Br$\gamma$, H$\alpha$, and V-K colors indicates
that the clusters show a considerable age spread from 4-13 Myrs.
The two star clusters in the interaction zone have ages near the lower
limit of that range, while those in the outer loops lie near the top of
that age range.  These clusters sample the overall age gradient visible
in the Antennae, the details of which we will discuss in a forthcoming
paper.  The equivalent widths of the CO band-heads indicate that the
starbursts in the nuclei are considerably older (65 Myrs; with the
nucleus of NGC 4038 having evidence for a significant contribution from
younger population with an age of $\sim$6 Myrs northward of its K-band
continuum peak).  The extinction is highly variable and especially high
in some parts of the overlap region (A$_V$=0.2 to 4.3).  Using these age
estimates and assuming the parameters of the IMF (Salpeter slope, upper
and lower mass cutoffs, etc.), we find that the clusters have masses
ranging between 0.3 to 3 $\times$ 10$^6$ M$_{\sun}$, (larger if the IMF
is extended below 1 M$_{\sun}$).  Our observations that the K-band
continuum peak (with its strong CO absorption) and the highest surface
brightness Br$\gamma$ emission are not coincident in some regions and
possible variations in W$_{CO}$ for clusters in [WS95]405 and the
nucleus of NGC 4038 argues for sequential star formation on small
scales.

\acknowledgments
We thank the staff of the AAT for their excellent assistance at the
telescope and the editor S. Willner and the referee for helping 
improve the presentation of our paper.

\clearpage

\clearpage

\begin{deluxetable}{llllllll}
\scriptsize
\tablecolumns{8}
\tablewidth{0pt}
\tablenum{1}
\tablecaption{Observed Properties of the Six Fields}
\tablehead{
\colhead{Property}&\colhead{Unit}&
\colhead{[WS95]86}&\colhead{[WS95]80}&
\colhead{[WS95]417}&\colhead{[WS95]405-A}&
\colhead{NGC 4039}&\colhead{NGC 4038}\\
\colhead{}&\colhead{}&\colhead{}&\colhead{}&
\colhead{[WS95]428}&\colhead{[WS95]405-B}&
\colhead{nucleus}&\colhead{A and B}\\
\colhead{(1)}&\colhead{(2)}&
\colhead{(3)}&\colhead{(4)}&
\colhead{(5)}&\colhead{(6)}&
\colhead{(7)}&\colhead{(8)}}

\startdata
R.A.		& J2000		& 12$^h01^m54^s.5$	& 12$^h01^m54^s.8$ & 12$^h01^m55^s.8$	
	    & 12$^h01^m50^s.6$	& 12$^h01^m53^s.4$ 	& 12$^h01^m53^s.0$ \\
Dec.		& J2000	& -18$^\circ$53\arcmin02\farcs2 & -18$^\circ$53\arcmin06\farcs0
			& -18$^\circ$52\arcmin10\farcs3 & -18$^\circ$52\arcmin11\farcs6
			& -18$^\circ$53\arcmin10\farcs3 & -18$^\circ$52\arcmin04\farcs3\\
Apert. size     &               & 2\farcs2$\times$2\farcs2 & 2\farcs2$\times$2\farcs2 
				& 1\farcs4$\times$1\farcs4 & 2\farcs2$\times$2\farcs2 
				& 3\farcs8$\times$3\farcs8 & 3\farcs8$\times$2\farcs6\\
		&		&		     	  & 	
				& 1\farcs4$\times$1\farcs0 & 2\farcs0$\times$1\farcs8
				&			   & 3\farcs0$\times$2\farcs2\\
W$_{Br\gamma}$	& \AA		& 210$\pm$10		& 116$\pm$10
				& 2$\pm$2 		& 10.0$\pm$2.5, & 2.9$\pm$0.9 & 3.2$\pm$0.9\\
		&	& &	& 7$\pm$3.0		& 54$\pm$5 & & 11$\pm$2 (58)\\
W$_{CO}$	& \AA	& 0$^{+2.0}_{-0.0}$ & 0$^{+2.0}_{-0.0}$ & 19$\pm$4 & 16.1$\pm$1.0 & 
10.2$\pm$0.9 & 10.0$\pm$0.9\\
		&	& & 	& 15$\pm$2 		& 0$\pm$2 &  & 8.3$\pm$0.9\\
Age		& 10$^6$yr	& 3.7$^{+1.0}_{-0.4}$  	& 5.5$^{+0.7}_{-0.8}$ 
				& 8.5...12.8 & 8.1$^{+2.0}_{-0.2}$ & 65$\pm$15 & 65$\pm$15\\
		& 		& 			& 	  & each & & & (5.9) \\
He I/Br$\gamma$  &               & 0.47$\pm$0.02         & 0.50$\pm$0.02 & & & & 0.17$\pm$0.03 \\
T$_{eff}$       & 10$^3$K       & 36$\pm$1              & 38$\pm$1	&    &  & & 33$\pm$1\\
m$_K$		& mag		& 14.5$\pm$0.2		& 14.8$\pm$0.2 
				& 15.5$\pm$0.3 		& 13.9$\pm$0.2 & 12.3$\pm$0.2 	& 12.9$\pm$0.2\\
	&	&	&	& each	 	&   		& 		& 13.3$\pm$0.2\\ 
m$_V$		& mag		& 16.3$\pm$0.15		& 18.8$\pm$0.15
				& 19.1$\pm$0.3, 	& 16.8$\pm$0.2 & 15.3$\pm$0.2 	& 16.1$\pm$0.2,\\ 
		&	&	& & 19.3$\pm$0.3	&   		&		& 16.6$\pm$0.2\\
A$_V$ 		& mag		& 1.4$\pm$0.3		& 4.3$\pm$0.3
				& 0.3$\pm$0.3		& 0.6$\pm$0.3
				& 0.2$\pm$0.1		& 0$^{+0.2}_{-0.0}$,\\
	& & & & 	& 	&			& 0.5$\pm$0.2\\
F$_{Br\gamma}$ 	& 10$^{-16}$ erg	& 83$\pm$8 		& 42$\pm$8 & 1.5$\pm$1 & 18$\pm$3
		& 20$\pm$1 		& 7$\pm$2 \\
		& s$^{-1}$cm$^{-2}$&		&	&	& (A+B) & & 21$\pm$3\\
F$_{H\alpha}$ 	& 10$^{-14}$ erg	& 32$\pm$3		& 2.3$\pm$0.2
				& 1.2$\pm$0.2 	& 10$\pm$1 & 17$\pm$1 & 8$\pm$1\\
		& s$^{-1}$cm$^{-2}$&	&	& (A+B)		& (A+B) & & 13$\pm$1\\
Mass		& 10$^6$M$_\sun$	& 1.6$^{+1.2}_{-0.2}$	& 3.0$^{+3.6}_{-0.7}$	& 0.4-0.6	& 1.1$\pm$0.3\\
		&		&		&		& each	& \\
\tablecomments{Col.  (1) --- Measured properties with 1$\sigma$
uncertainties.  Apert.  size is the size of the apertures from which the
values were derived.  W$_{Br\gamma}$ indicates the equivalent width of
the Br$\gamma$ line, W$_{CO}$ is the equivalent width of the $^{12}$CO
(2-0) bandhead feature, determined in the wavelength range 2.3060 -
2.3092 $\mu$m, which corresponds to the wavelength range given by
\citet{OO95}, redshifted to the radial velocity of the Antennae.  Its
uncertainty was estimated from the neighbouring continuum.  For
comparison, a few values of W$_{CO}$ for stars observed with 3D:  K4.5
Ib:  15.3\AA, M3.5 Iab:  19.6\AA, K3 III:  9.1\AA, M1 III:  12.1\AA\
\citep{Natasha99}.  Age is the model age (instantaneous burst,
Salpeter IMF between 1 and 100 M$_\odot$, solar metallicity) that most 
closely matches the
equivalent widths of Br$\gamma$ and CO.  Where the value for W$_{CO}$
covered a range that did not behave monotonically with age, we gave an
age range rather than a value $\pm\sigma$.  T$_{eff}$ of the most
massive stars as indicated by the He I/Br$\gamma$ ratio, He I/Br$\gamma$
measured from the 3D spectroscopy, m$_K$ from our spectroscopy, m$_V$
from HST images \citep{W99}, A$_V$ from the flux ratio of
Br$\gamma$/H$\alpha$, F$_{Br\gamma}$ as measured from our data,
F$_{H\alpha}$ from HST images \citep{W99}.  Col.  (2) --- Unit in which
the measurements in Cols.  (3) to (8) are tabulated.  Col.  (3) to (8)
--- The measured properties (given in Col.  (1)) of the six observed
fields.  The values for the properties of NGC 4038-B (Age, T$_{eff}$ and
the value for W$_{Br\gamma}$ in parentheses) make use of an assumption
about the continuum level, see text for details. Moreover, the age determination
of the nuclei may be influenced by the non-inclusion of the TP-AGB phase
in the Starburst99 models. }
\enddata
\end{deluxetable}

\end{document}